\documentclass[a4paper,11pt]{article}
\usepackage{pos}
\usepackage{siunitx}
\usepackage{multirow}
\usepackage{orcidlink}

\title{Next generation tracking and vertexing detectors}
\author*[a,1]{Armin Ilg\orcidlink{0000-0001-9488-8095}}

\note[1]{Now at Institut Pluridisciplinaire Hubert Curien (IPHC), Université de Strasbourg, CNRS/IN2P3, Strasbourg, France}

\renewcommand{\printHeadAuthors}{Armin Ilg}

\affiliation[a]{Physics institute, University of Zürich\\
  Winterthurerstrasse 190, 8057 Zürich, Switzerland}

\emailAdd{armin.ilg@cern.ch}

\abstract{Precise and efficient track and vertex reconstruction is essential to exploit the physics potential of collider experiments. The requirements of future tracking systems are determined by the collider type and its collision environment, required measurement precision, and beam structure. This contribution reviews the requirements and challenges of vertexing and tracking detectors at future $e^+e^-$, $e^-$-hadron, $\mu^+\mu^-$, and hadron colliders. 

Several common trends emerge across the different collider types. Tracking systems require increasingly precise spatial measurements, low-material designs, and, in some cases, integrated particle identification capabilities. Monolithic active pixel sensors are currently the leading option for vertex detectors at $e^+e^-$ and $e^-$-hadron colliders. Tracker concepts range from gaseous detectors, which provide many measurements per track and can achieve very low material budgets, to silicon and scintillating-fibre trackers, which provide fewer but more precise measurements and can tolerate higher hit rates. At \SI{10}{\tera\electronvolt} parton centre-of-mass colliders, precision timing throughout the tracking system is needed for beam-background rejection or pile-up mitigation. For future hadron colliders specifically, radiation tolerance beyond that demonstrated by existing sensor technologies is required.

Ultimately, meeting the vertexing and tracking requirements of future colliders will require substantial R\&D and the integrated optimisation of sensors, front-end electronics, readout, cooling, powering, and mechanics.}

\FullConference{14th Edition of the Large Hadron Collider Physics (LHCP2026)\\
18-22 May 2026\\
Paris, France\\}

\renewcommand{\logo}{\relax} % For arxiv only

\begin{document}

\maketitle

\section{Introduction}

Tracking systems are a central component of collider detectors. Their main goal is to reconstruct charged-particle tracks precisely and efficiently. The innermost tracking layers are commonly referred to collectively as the \textit{vertex detector} because their precise measurement of track impact parameters enables the reconstruction of primary, secondary and tertiary vertices.
%These vertices then are crucial for jet flavour tagging and 
The outer tracking layers form the \textit{tracker}. By measuring the curvature of charged-particle trajectories in the detector's magnetic field over a long lever arm, the tracker provides a precise momentum measurement.
%that is also input to particle-flow jet algorithms~\cite{Thomson2009}.

% For good tracking performance at current colliders, the individual sensors must provide a hit-detection efficiency of $\gtrsim 99\%$, good spatial resolution, and low noise. The sensors, support, readout, and cooling structures must also have a low material budget to minimise multiple Coulomb scattering. 

Key requirements include high hit-detection efficiency, low noise, and precise measurements of the incidence position of traversing particles. Further requirements include adequate timing to associate hits with the correct bunch crossing and track, sufficient radiation tolerance, and broad angular coverage. Future colliders must meet these requirements while also considering the detection of long-lived particles from the outset and the integration of particle identification (PID). 
%They may additionally require single-hit spatial resolutions of only a few $\SI{}{\micro\meter}$.

The properties of future collider tracking systems depend strongly on the collider type. The following factors are crucial:
\begin{itemize}
	\item \textbf{Collision environment}: Instantaneous luminosity and particle multiplicity determine the precision timing requirement (4D tracking), radiation tolerance, occupancy, and readout rate.
	\item \textbf{Precision}: The required accuracy of reconstructed tracks and vertices determines the requirements on spatial resolution and material budget.
	\item \textbf{Beam structure}: The bunch structure and triggering scheme determine the timing (bunch identification) and readout-rate requirements.
\end{itemize}

This contribution reviews $e^+ e^-$, $e^-$-hadron, $\mu^+\mu^-$, and hadron colliders, investigating how the collider type affects the requirements and designs of their vertexing and tracking detectors.

\section{Vertexing at $e^+ e^-$ colliders}

%Future $e^+e^-$ colliders all share similar features. 
Future $e^+e^-$ colliders have a mild to moderate collision environment, but demand the highest precision on their detectors. The beams are bunched, with either evenly spaced bunches or bunch trains being considered. The need for a hardware trigger depends on the machine and could even differ between experiments at the same collider. 

What is unique about vertex detectors at $e^+ e^-$ colliders is that they often are fully integrated into the machine-detector interface (MDI) region that also contains the %central and forward 
beam pipe, luminosity calorimeter and the first accelerator magnets. The minimal radius and maximal longitudinal extend of the vertex detector is thus constrained by the MDI. Being closest to the collisions, the vertex detector is also most affected by beam-induced backgrounds and in general the beam properties. The dominant backgrounds are incoherent pair creation and synchrotron radiation that drive the hit rates and timing requirements. 
%At circular $e^+e^-$ colliders, bunch collisions are currently foreseen at $\mathcal{O}(\SI{40}{\mega\hertz})$, while 
Linear colliders feature bunch trains with tightly packed bunches and long periods without bunches in between, allowing to turn off detector components in-between (\textit{power pulsing}). This reduces the detector power consumption in linear colliders and thus simplifies cooling. In both linear and circular $e^+ e^-$ collider vertex detector designs, air-cooling is the baseline for the first couple of layers to minimise material budget.

The vertex detector performance requirement can be stated as a requirement to the transverse impact parameter resolution, which is commonly parametrised as $\sigma_{d_0} = a \oplus \frac{b}{p \sin^{3/2} \theta}$. $a$ is given by the sensor resolution and distance to the interaction point. Thus, pixel sensors are needed and should be placed as close as possible. $b$ describes the effect of multiple Coulomb scattering, thus the material budget in the vertex detector and the beam pipe in front must be minimised. For FCC-ee, $a$ of about \SI{3}{\micro\meter} and $b$ of \SI{15}{\micro\meter\giga\electronvolt} are currently targeted~\cite{FCC:2025lpp}, but better values would, for example, improve the measurement of the $B^{0} \rightarrow K^{*0} \tau^+ \tau^-$~\cite{Miralles_2024Ks}. The $b$ requirement can be a bit relaxed at higher-$\sqrt{s}$ machines due to the dependence on the particle momentum.

The only sensor technology currently considered for future $e^+ e^-$ vertex detectors are monolithic active pixel sensors (MAPS). They integrate the amplification, and readout into the same silicon die as the sensor, thus allowing for the construction of very light detectors. Thanks to advanced CMOS processes, small pixels of $\sim 20\times\SI{20}{\micro\meter\squared}$ can be manufactured, while keeping power consumption low and timing performance reasonable. Table~\ref{tab:MAPS} lists the properties or specifications of three MAPS R\&D lines targeting future $e^+ e^-$ collider vertex detectors.

\begin{table}[htbp]
	\centering
	\footnotesize
	\begin{tabular}{p{0.215\linewidth}|p{0.215\linewidth}|p{0.275\linewidth}|p{0.185\linewidth}}

		                                                             &
		\textbf{ARCADIA~\cite{Pancheri2020,DaRochaRolo2025}}         &
		\textbf{OCTOPUS~\cite{King:2026ocl}}                         &
		\textbf{TaichuPix~\cite{Wu:2023ann}}                                                                                                   \\ \hline

		Process                                                      &
		\SI{110}{nm} LFoundry                                        &
		\SI{65}{nm} TPSCo                                            &
		$\SI{180}{\nano\meter}$ TJ                                                                                                             \\ \hline

		Pitch [\SI{}{\micro\meter\squared}]                          &
		$25\times25$                                                 &
		$\mathcal{O}(20\times20)$                                    &
		$25\times25$                                                                                                                           \\ \hline

		$\sigma_{\text{xy}}$[\SI{}{\micro\meter}]                    &
		Down to $4.6$~\cite{Pantouvakis:2026qkh}                     &
		Goal: $3$                                                    &
		Down to $4.5$~\cite{Li:2024mhw}                                                                                                        \\ \hline

		Timing                                                       &
		$\mathcal{O}(\SI{}{ns})$ to $\mathcal{O}(\text{10's of ps})$ &
		$\SI{5}{\nano\second}$ time tag with ToT                     &
		$< \SI{100}{ns}$ time walk~\cite{Wang:2023ppi} \\ \hline

		Size [\SI{}{\centi\meter\squared}]                           &
		$1.28\times1.28$, side-abuttable                             &
		$\approx 2\times3$                                           &
		$2.56\times1.28$                                                                                                                       \\ \hline

		Thickness [\SI{}{\micro\meter}]                              &
		$< 50$ to 500                                                &
		$\leq 50$                                                           &
		150                                                                                                                                    \\ \hline

		Power [\SI{}{\milli\watt\per\centi\meter\squared}]           &
		$< 30$                                                       &
		$\leq 50$ (goal), 500 (first prototype)	          &
		$< 200$                                                                                                                                \\ \hline

		Pixel hit rate [\SI{}{\mega\hertz\per\centi\meter\squared}]  &
		Up to 100                                                    &
		100       &
		Spec: 36~\cite{Zhang:2022rlo}                                                                                                          \\ \hline

		Project status                                               &
		Full-scale prototype                                 &
		Design phase                                                 &
		Full-scale prototype                                                                                                \\
	\end{tabular}
	\caption{Selection of MAPS R\&D lines towards future $e^+ e^-$ vertex detectors.}
	\label{tab:MAPS}
\end{table}

Vertex-detector layouts studied for FCC-ee and CEPC include stave-based geometries and ultra-light, wafer-scale bent MAPS~\cite{Ilg2024,CEPCStudyGroup:2025kmw}. The novel FCC-SEED concept foresees overlapping, curved MAPS layers, with power and readout signals routed through a flexible circuit.

\section{Tracking at $e^+ e^-$ colliders}

% The main deliverable of trackers at $e^+ e^-$ colliders is an exquisite charged-particle momentum resolution. 
Precise track reconstruction is crucial for much of the $e^+ e^-$ collider physics programme with the prime example being the measurement of the Higgs boson mass.
% A prime example is the Higgs mass measurement in the $Z(\mu^+\mu^-)H$ channel, where the fundamental limit set by the beam-energy spread can only be reached with excellent muon momentum resolution.
%Precisely knowing the particle momenta is also a crucial input to particle flow jet reconstruction and thus the jet energy resolution. 
At the FCC-ee, the consevative (aggressive) requirement on the charged-particle momentum resolution is $\sigma_p/p < 0.2\%$ ($< 0.1\%$) for tracks with momenta of $\mathcal{O}(\SI{50}{\giga\electronvolt})$. Achieving such a resolution requires a minimisation of the material budget (esp. at low-$\sqrt{s}$ colliders) and the hit resolution (especially at high-$\sqrt{s}$ colliders).

Table~\ref{tab:trackers} summarises the tracking technologies currently considered for high-energy future colliders. Two approaches are being pursued: a small number of layers providing highly precise spatial measurements, as in silicon and scintillating-fibre (SciFi) trackers, or gaseous trackers that combine many less precise hits to achieve good overall resolution, as in drift chambers, time-projection chambers (TPCs), and straw tubes. Gaseous trackers offer low material budgets and continuous tracking, which can benefit searches for long-lived particles. Silicon and SciFi trackers can tolerate higher hit rates, which can be challenging for gaseous trackers with long integration times, such as drift chambers, or charge accumulation from beam-induced backgrounds, as in TPCs.

\begin{table}[htbp]
    \centering
    \footnotesize

    \begin{tabular}{p{0.132\textwidth}|p{0.135\textwidth}|p{0.15\textwidth}|p{0.13\textwidth}|p{0.135\textwidth}|p{0.135\textwidth}}
            & \textbf{Drift chamber}
            & \textbf{TPC}
            & \textbf{Straw tubes}
            & \textbf{Silicon tracker}
			& \textbf{SciFi} \\ \hline

        Proposed for detector concept
            & IDEA, ALLEGRO
            & ILD, AGORA
            & ALLEGRO
            & CLD, CLICdet, SiD, ALFA, MAIA, MUSIC
			& ALLEGRO \\

        Hit count
            & $\mathcal{O}(100)$
            & $\mathcal{O}(100)$
            & $\mathcal{O}(100)$
            & $\mathcal{O}(5$--$10)$
			& $\mathcal{O}(5$--$10)$ \\

        Hit resolution
            & $\sim100\,\mu\mathrm{m}$
            & $\sim100\,\mu\mathrm{m}$
            & $\sim100\,\mu\mathrm{m}$
            & Few \SI{}{\micro\meter} 
			& Tens of \SI{}{\micro\meter} \\
        Material budget
            & Ultra-low
            & Low
            & Low
            & Medium 
			& Low-medium \\

        PID capability
            & $\mathrm{d}N/\mathrm{d}x$
            & $\mathrm{d}E/\mathrm{d}x$
            & $\mathrm{d}E/\mathrm{d}x$
            & $\mathrm{d}E/\mathrm{d}x$ (limited) or time-of-flight
			& Time-of-flight (limited) \\

        Concerns and challenges
            & Wire tension, wire failure
            & Field distortions from ion backflow
            & Stereo angle, mechanics
            & Cooling and services 
			& Light yield along long fibre \\
	\end{tabular}
	\caption{Overview of tracker technologies for future high-energy colliders.}
	\label{tab:trackers}
\end{table}

At circular colliders, the detector magnetic field is limited to \SI{3}{\tesla} to minimise its impact on the beams and preserve luminosity\footnote{Working hypothesis, previously limited to \SI{2}{\tesla}.}. Large trackers, with $r_\text{max} \sim \SI{2}{m}$, compensate for this limitation. At linear $e^+ e^-$ colliders, magnetic fields can be substantially higher\footnote{\SI{5}{\tesla} in the case of SiD~\cite{Breidenbach:2021sdo}.}, allowing for smaller trackers.
% The tracker radius can therefore be reduced and instead the calorimeters are deeper to contain the more energetic showers produced at high-$\sqrt{s}$ linear colliders.

\section{The ePIC tracking system at the Electron-Ion Collider}

The Electron-Ion Collider (EIC~\cite{Abdul_Khalek_2022}) will operate at $\sqrt{s}$ up to \SI{140}{\giga\electronvolt} and an instantaneous luminosity of \SI{1e34}{\per\centi\meter\squared\per\second}. Its collision environment and reconstruction requirements are less demanding than those of the $e^+ e^-$ colliders discussed above\footnote{High-energy, high-intensity electron-hadron colliders such as the LHeC or FCC-eh would have harsher radiation environments and higher hit rates than the EIC.}. The EIC therefore provides an intermediate case between (HL-)LHC trackers and trackers for future colliders.
%, both in terms of requirements and in timescale.

The ePIC detector at the EIC features a \SI{1.7}{\tesla} magnetic field and a tracker comprising three complementary subsystems~\cite{turrisi2026epicsiliconvertextracker}: a vertex detector with a \SI{2}{\micro\second} integration time, a moderately fast micro-pattern gas detector (MPGD) with $\sigma_\text{t} \approx \SI{10}{ns}$, and a fast time-of-flight (TOF) system with $\sigma_\text{t} \approx \SI{30}{ps}$. The three inner barrel layers of the vertex detector use MOSAIX, wafer-scale bent MAPS developed in the ALICE ITS3\cite{its3_tdr} R\&D programme in the TPSCo \SI{65}{\nano\meter} process. The two outer barrel layers and five disks use a flat, adapted version of MOSAIX. The MPGD comprises one inner and one outer barrel layer and two disks on each side; its timing information supports pattern recognition and complements the slower vertex detector. Finally, the TOF system consists of a barrel between the MPGDs and a forward disk, both using AC-coupled low-gain avalanche detectors (AC-LGADs) for PID.

The ePIC tracking system illustrates the benefits of combining technologies optimised for different tasks, although this approach requires multiple sensors, ASICs, and readout systems to be developed in parallel. An alternative is to develop versatile sensors that can be tuned to different requirements. For example, the DRD3 MANTA project aims to cover the timing and power-consumption requirements of detectors ranging from the ALICE3, Belle II, and LHCb upgrades to CBM, and the FCC-ee tracker and potentially timing layer.

Another approach is to combine multiple functions in a single sensor, as pursued by the CASSIA~\cite{Haberl:2026eqy} and ARCADIA MADPix~\cite{Follo:2024zej} R\&D programmes. These projects develop MAPS with a gain layer to achieve both good spatial and timing performance.

% Finally, looking at high-energy, high-intensity electron-hadron colliders like the LHeC or FCC-eh, these would have a harsher radiation environments and higher hit rates than the EIC.

\section{Vertexing and tracking at $\mu^+ \mu^-$ colliders}

At $\mu^+\mu^-$ colliders, the collision environment is much harsher than at the collider types discussed above. At beam energies of up to \SI{5}{\tera\electronvolt}, muon decays produce a large background of hits in the detector. This requires a complex MDI design and timing resolutions of tens of picoseconds in all tracking layers to suppress these background hits. The associated total ionising dose and non-ionising energy loss 
in the vertex detector and tracker 
can be comparable to those at the HL-LHC~\cite{InternationalMuonCollider:2025sys}. The high-momentum particles produced at collision energies up to $\sqrt{s} = \SI{10}{\tera\electronvolt}$ also impose stringent spatial-resolution requirements. For example, the MAIA detector concept~\cite{MAIA:2025hzm} assumes spatial resolutions of $5\times \SI{5}{\micro\meter\squared}$ in the vertex detector and $7\times \SI{90}{\micro\meter\squared}$ in the inner tracker.

The low collision rate from the single $\mu^+$ and $\mu^-$ bunches may allow every collision to be analysed and enable power pulsing to reduce sensor power consumption and cooling requirements.

% \begin{table}[htbp]
% 	\centering
% 	\begin{tabular}{l|c|c|c}
% 								& Vertex Detector
% 								& Inner Tracker
% 								& Outer Tracker \\
% 		\hline

% 		Sensor type                & pixels
% 								& macro-pixels
% 								& micro-strips \\

% 		Barrel Layers              & 4
% 								& 3
% 								& 3 \\

% 		Endcap Layers (per side)   & 4
% 								& 7
% 								& 4 \\

% 		Cell Size                  & $25\,\mu\mathrm{m} \times 25\,\mu\mathrm{m}$
% 								& $50\,\mu\mathrm{m} \times 1\,\mathrm{mm}$
% 								& $50\,\mu\mathrm{m} \times 10\,\mathrm{mm}$ \\

% 		Sensor Thickness           & $50\,\mu\mathrm{m}$
% 								& $100\,\mu\mathrm{m}$
% 								& $100\,\mu\mathrm{m}$ \\

% 		Time Resolution            & $30\,\mathrm{ps}$
% 								& $60\,\mathrm{ps}$
% 								& $60\,\mathrm{ps}$ \\

% 		Spatial Resolution         & $5\,\mu\mathrm{m} \times 5\,\mu\mathrm{m}$
% 								& $7\,\mu\mathrm{m} \times 90\,\mu\mathrm{m}$
% 								& $7\,\mu\mathrm{m} \times 90\,\mu\mathrm{m}$ \\
% 	\end{tabular}
% 	\caption{Performance assumptions of the vertex and tracker sensors in the MAIA detector concept. Reproduced from ~\cite{MAIA:2025hzm}.}
% 	\label{tab:MAIA}
% \end{table}

\section{Vertexing and tracking at future hadron colliders}

Finally, future hadron colliders such as the FCC-hh or SPPC present the harshest collision environments, with a pile-up of up to 1000 and a centre-of-mass energy of $\sqrt{s}=\mathcal{O}(\SI{100}{\tera\electronvolt})$. The radiation levels are expected to be 1--2 orders of magnitude higher than those at the HL-LHC, reaching up to $10^{18}\ \text{n}_\text{eq}/\si{\centi\meter\squared}$ and \SI{300}{\mega\gray}~\cite{FCC:2018vvp}. Resolving hard-scatter collisions amid the pile-up requires a time resolution of $\lesssim \SI{30}{\pico\second}$ in all tracking layers, enabling 4D tracking. The relevant track momenta range from \SI{20}{\giga\electronvolt} to \SI{20}{\tera\electronvolt}. To achieve a momentum resolution of 20\% at \SI{10}{\tera\electronvolt} while keeping the occupancy below 1\%, granularities as small as $25 \times \SI{50}{\micro\meter\squared}$ are required.
 %, depending on the layer. 
In the FCC-hh reference detector, the tracker outer radius is currently assumed to be \SI{1.6}{\meter}~\cite{FCC:2018vvp}, substantially larger than the \SI{1}{\meter} (\SI{1.1}{\meter}) of ATLAS (CMS) and therefore requiring a larger volume to be instrumented. In addition, next-generation hadron-collider trackers must provide extended forward coverage up to $|\eta| < 6$, which necessitates dedicated forward tracking systems.

No sensor technology exists yet with adequate performance and radiation tolerance. Relevant R\&D programmes include 3D sensors, LGADs, and wide-band-gap materials~\cite{Cheng2025}.

\section{Conclusions}

Vertexing and tracking detectors are central to all current and future collider experiments. Their requirements are driven by the collision environment, the required measurement precision, and the beam structure, which in turn depend on the collider type, $\sqrt{s}$, and instantaneous luminosity. The examples discussed here span a broad range of requirements, from the precision-dominated $e^+e^-$ colliders to the intense backgrounds and radiation fields of $\mu^+\mu^-$ and hadron colliders. The EIC occupies an intermediate position and demonstrates how instrumentation R\&D developed for the HL-LHC can support future collider trackers. Across all collider types, trackers must be designed as integrated systems, with powering, DAQ, mechanics, and cooling optimised alongside sensor and front-end performance from the outset to fully realise the physics potential of future colliders.

\newpage

\acknowledgments

The author thanks the Swiss National Science Foundation (SNSF) for its support under Grant number 223515. Generative AI tools (Microsoft Copilot) were used to improve the language and grammar of this manuscript. The content, ideas, and mistakes are the author's own.

\bibliographystyle{JHEP}
\bibliography{biblio}

\end{document}